\documentstyle[amssymb,aps,prl]{revtex}

\begin{document}

\twocolumn[\hsize\textwidth\columnwidth\hsize\csname
@twocolumnfalse\endcsname

\draft
\author{Gonzalo Usaj$^1$, Horacio M. Pastawski$^{1,2}$, and
Patricia R. Levstein$^1$}
\address{$^{1}$Facultad de Matem\'{a}tica, Astronom\'{\i }a y F\'{\i }sica,
Universidad Nacional de C\'{o}rdoba, Ciudad Universitaria, 5000 C\'{o}rdoba,Argentina}
\address{$^{2}$International Center for Theoretical Physics, P. O. Box 586, 34100,Trieste, Italy}
\title{Gaussian to Exponential Crossover in the Attenuation of Polarization
Echoes in NMR}
\date{Published in Molecular Physics {\bf 95}(6) 1229 (1998)}
\maketitle

\begin{abstract}
An ingenious pulse sequence devised by S. Zhang , B. H. Meier , and R. R.
Ernst (Phys. Rev. Lett. {\bf 69}, 2149 (1992)) reverses the time evolution
(``spin diffusion'') of the local polarization in a dipolar coupled $^{1}$H
spin system. This refocusing originates a Polarization Echo whose amplitude
attenuates by increasing the time $t_R$ elapsed until the dynamics
is reversed. Different functional attenuations are found for a set of
 dipolar
coupled systems: ferrocene, (C$_5$H$_5$)$_2$Fe, cymantrene, 
(C$_5$H$_5$)Mn(CO)$_3$, and cobaltocene, (C$_5$H$_5$)$_2$Co. 
To control a relevant variable involved in this attenuation a pulse sequence
has been devised to progressively reduce the dipolar dynamics. Since
 it
reduces the evolution of the polarization echo it
 is referred
as REPE sequence. Two extreme behaviors were found while
characterizing the materials: In systems with a strong source of
relaxation and slow dynamics, the attenuation follows an exponential law
(cymantrene). In systems with a strong dipolar dynamics the attenuation is
mainly Gaussian. By the application of the REPE sequence the characteristic
time of the Gaussian decay is increased until the presence of an underlying
dissipative mechanism is revealed (cobaltocene). For ferrocene, however, the
attenuation remains Gaussian within the experimental time scale. These two
behaviors suggest that the many body quantum dynamics presents an extreme
intrinsic instability which, in the presence of small perturbations, leads
 to
the onset of irreversibility. This experimental conclusion is consistent
 with
the tendencies displayed by the numerical solutions of model systems.
 
\end{abstract}
 \bigskip
] \narrowtext

\section{Introduction}

Nuclear spin dynamics, often called spin diffusion, is a powerful
tool\cite
 {NMR} to analyze both the local and long range structure in solids.
Since
 the spin diffusion rate depends on the internuclear dipolar
interactions, it
 contains useful information on the spatial proximity of the
nuclei and on
 the dimensionality of the interaction network\cite{ZENO}.
Therefore, there
 is a great interest in new pulse sequences applicable to a
broad range of
 systems. This is the case of the pulse sequence devised by S.
Zhang, B. H.
 Meier, and R. R. Ernst\cite{ZME} (ZME) which creates a {\em local}
polarization (LP) in a homonuclear spin system ($I$-spins) and detects
its
 later evolution. A central point of this sequence is that it uses the
presence of rare $S$-spins to label each of the $I$-spins directly bonded to
them. Therefore, no spectral resolution of $I$-spins is required. This
sequence has been successfully applied to several systems with different
purposes\cite{ZME2,Hirs,PUL,LUP} and, recently, it has been generalized\cite
{Tomaselli} to be used under magic angle spinning (MAS) increasing its
potential applications. On the other hand, it allows one to monitor the
formation of a {\em polarization echo} (PE) \cite{ZME} when, by external
means, the LP evolution is reversed. This is achieved by a change of the
sign of the effective dipolar Hamiltonian \cite{Rhim}. The initial localized
polarization evolves into multiple-spin order during a time $t_{R}$ and
then, suitable pulses switch ${\cal H}\rightarrow -[2]{\cal H}$, retracing
the dynamics and building up a PE at $t\thickapprox t_{R}+\left[ \frac{1}{2}%
\right] t_{R}$.

Despite the fact that the dominant part of the Hamiltonian is reversed, the
PE amplitude ($M_{PE}$) attenuates as the time $t_{R}$\ grows. The study of
this attenuation might provide a new characterization tool which so far has
not been exploited. Furthermore, it should be possible to use the PE\
attenuation to obtain additional information on the underlying stochastic
processes that destroy the dipolar spin coherence and inhibit the echo
formation. Such procedure resembles the use of the Hahn's echoes\cite{Hahn}
to detect molecular diffusion\cite{CPMG}. In this sense, there is an obvious
interest in experimental approaches allowing a proper characterization of
this attenuation for different dipolar coupled systems. While a dependence
of $M_{PE}$\ on time and the spinning rate has been already reported\cite
{Tomaselli}, there is still a need to extend these studies to different
systems. As a first step to use the PE attenuation as a practical tool, we
should investigate its functional dependence on $t_{R}$\ and, if possible,
identify its physical origin. In principle, there could be many possible
mechanisms contributing to the attenuation of the PE. A non-exhaustive list
includes: non inverted terms in the Hamiltonian, pulse imperfections, local
relaxation, spin motion, etc. Our idea involves the use of a set of systems
where most of these variables remain essentially unchanged. Within the
metallocenes family we can find compounds which allow us to change the
dipolar interaction network and independently the magnetic nature of the
metallic nuclei, this is, we can change the strength of a local source of
relaxation. This would allow us to analyze whether the dipolar interaction
plays any role in the attenuation and how the presence of irreversible
processes manifests in the decay rate.

In this work, we used the ZME sequence to study the PE attenuation in a set
of polycrystalline samples. In all these systems, the molecular structure
contains one or two cyclopentadienil rings where only one $^{1}$H-spin is
labeled by the rare $^{13}$C-spin. Our results confirm, as it was suggested
by recent experiments\cite{LUP}, that there are systems where the decay is
exponential while there are others where it is Gaussian. However, a
combination of both behaviors is observed in a third class of systems. In
order to understand this and, at the same time, to test the role of the
dipolar interaction, we devised an {\it ad hoc} multiple pulse sequence,
based on the ZME one, to {\em progressively }slow down the dipolar dynamics
while the non-invertible interactions are kept constant.

While other mechanisms cannot be completely ruled out, our experimental
results are consistent with the hypothesis that the dipolar dynamics
controls the attenuation. In some samples, if the dynamics is sufficiently
reduced, a crossover from the Gaussian to the exponential attenuation law
can be observed. We present exact numerical solutions of the spin dynamics
in a cyclopentadienil ring and in a double ring molecule, which give further
support to our hypothesis. They illustrate how the spin dynamics, in the
presence of either small residual interactions or strong relaxation
processes, increases the PE attenuation.\ 

\section{Experimental Methods}

We analyzed the PE attenuation in polycrystalline samples of ferrocene\cite
{Seiler}, (C$_{5}$H$_{5}$)$_{2}$Fe, cymantrene\cite{Fizpatrick}, (C$_{5}$H$%
_{5}$)Mn(CO)$_{3}$, and cobaltocene\cite{Antipin}, (C$_{5}$H$_{5}$)$_{2}$Co
and in a single crystal sample of ferrocene, all with natural $^{13}$C
isotopic abundance. At room temperature, these compounds crystallize in a
monoclinic form with space group $P2_{1}/a$. The cyclopentadienil rings
perform fast rotations around their five-fold symmetry axis leading to
inhomogeneous $^{13}$C spectra, where the resonance frequency depends on the
angle between the external magnetic field and the molecular symmetry axis.
All the NMR measurement were performed in a Bruker MSL-300 spectrometer,
equipped with a standard Bruker CP-MAS probe.

The ZME sequence to reverse the LP evolution is schematized in Fig. 1a. As
we mentioned above, the central idea is to use the rare $^{13}$C spin ($S$%
-spin) as a {\em local} probe to inject magnetization to one of the abundant 
$^{1}$H-spins ($I$-spins) and to capture what is left after the $I$-spins
have evolved. In the first part, the $S$-spin is polarized by means of a
Hartmann-Hahn cross polarization\cite{HH} (CP). After that, the $S$
polarization is kept spin-locked for a time $t_{S}$ while the remaining
proton coherence decays to zero. Then, {\bf A) }A short CP pulse of duration 
$t_{d}$ selectively repolarize the $I_{1}$-spin directly bonded to the $S$%
-spin, creating the initial localized state ($t_{d}$ is the time when the
first maximum in a simple cross polarization transfer occurs\cite{Muller}); 
{\bf B) }The LP evolves during a time $t_{1}$ in the rotating frame under a
strong spin-lock field with the effective Hamiltonian 
\begin{equation}
{\cal H}^{y}=\left[ -\frac{1}{2}\right] \sum_{j>k}\sum_{k}d_{jk}\,\left[
2I_{j}^{y}I_{k}^{y}-\frac{1}{2}\left(
I_{j}^{+}I_{k}^{-}+I_{j}^{-}I_{k}^{+}\,\right) \right] .  \label{Hii}
\end{equation}
The interaction parameters 
\begin{equation}
d_{jk}=-\frac{\mu _{0\,\,}\gamma _{I}^{2}\,\,\,\,\hbar ^{2}}{4\pi r_{jk}^{3}}%
\,\,\frac{1}{2}\left\langle 3\cos ^{2}\theta _{jk}-1\right\rangle ,
\end{equation}
are time averaged due to the fast rotation of the rings. Here, $\gamma
_{I}^{{}}$ is the gyromagnetic factor of the $I$-spin, the $r_{jk}^{{}}$'s
are their internuclear distances and $\theta _{jk\text{ }}$are the angles
between internuclear vectors and the static magnetic field. {\bf C}){\bf \ }%
A $\left( \pi /2\right) _{x}$ pulse tilts the polarization to the laboratory
frame where the $I$-spins evolve with $-\left[ 2\right] {\cal H}^{z}$ during
a time $t_{2}$. $\,$A $\left( \pi /2\right) _{-x}$ pulse leads the
polarization back to the rotating frame. During $t_{2}$, the effective
Hamiltonian $-\left[ 2\right] {\cal H}^{y}$, produces the refocusing that
builds up the PE at $t_{2}\thickapprox \left[ \frac{1}{2}\right] \left(
t_{1}+t_{d}\right) =\left[ \frac{1}{2}\right] t_{R}$\cite{PUL}. {\bf D)} A
short CP pulse transfers the polarization back to $S$. {\bf E) }The $S$
polarization is detected while the protons are kept irradiated. The PE
attenuation can be monitored as a function of $t_{R}$ by setting different
values for $t_{1}$. As discussed below, a sequence where the evolution in
the laboratory frame precedes the refocusing in the rotating frame (Fig. 1b)
is more appropriate to compare attenuations in different systems.

In order to test the effect of the dipolar dynamics in the PE attenuation,
we replace the {\bf B}) and {\bf C}) parts of the ZME sequence by $n$
defocusing and refocusing periods of $t_{1}=t_{R}/n$ and $t_{2}=\left[ \frac{%
1}{2}\right] t_{1}$ respectively. Therefore, the transfer of single-spin
order into multiple-spin order, for a given time $t_{R}$, can be gradually
reduced by increasing $n$. Hereafter we will refer to this as the Reduced
Evolution Polarization Echo (REPE) sequence. It includes extra refocusing
periods $t_{m}/2\approx \left[ \frac{1}{2}\right] t_{d}/2$ to compensate the
proton spin evolution during the CP periods. This allows us to have an
experimental point to properly normalize the data. The resulting multiple
pulse sequence is shown in Fig. 1c. For $n=1$ the REPE sequence differs from
the ZME one by the extra periods $t_{m}/2$. From a practical point of view,
it is important to adjust the relative phase and the RF amplitude of the $X$,%
$-X$,$Y$,$-Y$ channels\cite{Burum} since small errors produce a significant
signal loss.

\section{Results and discussion}

As a first step, we measured the PE amplitude as a function of $t_R$ for all
polycrystalline samples. The normalized experimental data are shown in Fig.
2. These experimental data correspond to molecules with their symmetry axes
approximately perpendicular to the external magnetic field (they were
frequency selected from the $^{13}$C spectrum). We used the ZME sequence
with $\omega _{1I}/2\pi =44.6$ {\rm kHz}, $t_C=2$ $m{\rm s}$, $t_S=1$ $m{\rm %
s}$, $t_d=85$ $\mu {\rm s}$, for ferrocene and $t_d=95$ $\mu {\rm s}$, for
cymantrene. In the case of cobaltocene, the short relaxation time $T_{1\rho
}^S\approx 780$ $\mu {\rm s}$ causes a significant $^{13}$C signal loss if
long CP or spin lock periods are used. Therefore, we maximized the first
polarization transfer by setting $t_C=t_d=85$ $\mu {\rm s}$ and we chose $%
t_S=150$ $\mu {\rm s}$ which is long enough to allow for the loss of proton
spin coherence and short enough to prevent a considerable signal loss.

Clearly, the three samples show a different functional dependence of the
decay of the PE on $t_{R}$. In the cymantrene sample the decay is clearly
exponential while in ferrocene it is Gaussian. The decay for cobaltocene is
more complex. Its fitting was possible only after the implementation of the
pulse sequence of Fig. 1c as explained below. It contains both exponential
and Gaussian contributions. Taking into account that in ferrocene the
intermolecular $I$-$I$ interactions are very important while in cymantrene
the rings stay relatively isolated within the experimental time scale, we
set asymptotic values of $0$ and $0.2$ respectively. Thus, the solid lines
are fittings with the characteristic time as the only free parameter. In the
case of cobaltocene there are two characteristic times.

The exponential decay can be considered as a manifestation of the presence
of a strong irreversible process characterized by a time $\tau _{\phi }$.
According to the Fermi Golden rule, the exponential decay is the signature
of a weak interaction with each of the states in a wide band of a continuous
spectrum\cite{AFA98}. The interaction with each degree of freedom is weak
but the overall effect results in a strong irreversible process. What
interactions are present in cymantrene and cobaltocene but not in ferrocene?
In the cymantrene sample, the strong local relaxation could be due to the
quadrupolar nature of the Mn nucleus in a low symmetry environment while in
cobaltocene it is more probably due to the paramagnetic nature of the Co(II).

The sources of Gaussian decay in ferrocene and the Gaussian factor in
cobaltocene are less obvious. When we take into account the magnitude of the
intermolecular dipolar interaction in these compounds, we observe that it
grows from cymantrene to cobaltocene. Then it is important to evaluate if
the magnitude of the dipolar interaction relative to $\hbar /\tau _{\phi }$
is playing any role in determining the type of decay.

It should be mentioned a subtle issue underlying the comparison of the
attenuations in different systems. Since there is some $I$-spin evolution
during the cross polarization time, we do not have an experimental value for 
$M_{PE}$ at $t_{R}=0$. Hence, the normalization of the signals depends on
the functional decay chosen in the fittings. While in ferrocene and
cymantrene systems this was not particularly demanding because of the
obvious simple functional dependences they presented, this can be a serious
restriction in a more general situation. In such a case, the pulse sequence
sketched in Fig. 1b should be used. Here, the amplitude at the maximum $%
M_{PE}(t_{2}=t_{m};t_{1}=0)$ constitutes an experimental reference to
normalize regardless of the functional law of the decay. The other way to
overcome this problem is to employ the REPE sequence sketched in Fig. 1c and
described in the experimental section. This has the additional advantage of
controlling the complexity reached by the dipolar order.

We applied the REPE sequence to a ferrocene single crystal. Fig. 3 shows the 
$^{13}$C-NMR spectrum (with the sequence parameters given in the caption).
Each peak corresponds to one of the two magnetically inequivalent sites in
the crystal. The carrier frequency was finely tuned to correspond to one of
them. The good signal to noise ratio allows us to detect small fractions of
the initial polarization. The attenuation of the PE for $n=1$ is shown in
Fig. 4. The well-defined Gaussian decay confirms the results obtained for
the polycrystalline sample. The compensating periods $t_{m}/2$ allow us to
properly normalize the data at $t_{R}=0.$ Besides, the unbounded nature of
the spin network justify an asymptotic value of $0$ for $M_{PE}$. The solid
line represents a fitting with $M_{PE}(t_{R})=\exp (-\frac{1}{2}(t_{R}/\tau
_{mb}^{1})^{2})$, obtaining $\tau _{mb}^{1}=(245\pm 5)$ $\mu {\rm s}$. When
we set $n=2,8,16$\ (inset Fig. 4), the Gaussian decay is kept but its
characteristic time $\tau _{mb}^{n}$\ grows. As we discuss below, this is in
agreement with our hypothesis that the time scale for the Gaussian
attenuation is controlled by the dipolar interaction itself. We found that $%
\tau _{mb}^{n}$\ grows as $\tau _{mb}^{n}=a\cdot n+b$\ with $a=\left( 54\pm
2\right) $\ $\mu {\rm s}$\ and $b=\left( 210\pm 10\right) \,\mu {\rm s}$.

When applying the same sequence to the cobaltocene sample we observe a
rather different behavior. While the dipolar interaction network is quite
similar to the ferrocene one, the paramagnetic Co(II) atom introduces a
strong source of relaxation. Thus, we expect a dipolar dynamics similar to
that of ferrocene but convoluted with an extra relaxation mechanism. Fig. 5
shows the PE attenuation in the cobaltocene sample as a function of $t_{R}$
for $n=1,2,5,8,16$. There is a clear crossover from a Gaussian decay to an
exponential one. This is the main result of this paper. The solid lines are
fittings of the experimental data to the equation 
\begin{equation}
M_{PE}(t_{R})=\exp \left[ -t_{R}/\tau _{\phi }-\frac{1}{2}\left( t_{R}/\tau
_{mb}^{n}\right) ^{2}\right] .  \label{Mpe}
\end{equation}
Based on the results for ferrocene we set a linear dependence of $\tau
_{mb}^{n}$ on $n$. This leaves only three free parameters for the whole data
set. By increasing $n,$ $M_{PE}$ reaches an asymptotic exponential decay
with $\tau _{\phi }=(640\pm 20)$ $\mu {\rm s}$. This shows that the Gaussian
factor can be gradually reduced until the underlying exponential decay
becomes dominant. Once this regime is reached, further increment of $n$ does
not produce any effect. Expression (\ref{Mpe}) must be considered as an
empirical one to account for the experimental results.

At this point it is important to mention the difference between our results
and those that could be expected by considering the analogies of the REPE
sequence with the Carr-Purcell-Meiboom-Gill (CPMG) pulse sequence\cite{CPMG}
used to identify the contributions that attenuate the Hahn echo. If a single
echo pulse sequence is applied, the amplitude of the Hahn echo at time $%
t_{H} $ is proportional to $\exp [-t_{H}/T_{2}-(t_{H}/\tau _{D}^{{}})^{3}]$.
Here $T_{2}$\ represents some processes not inverted by the Hahn echo and $%
\tau _{D}^{{}}$\ the molecular diffusion process in presence of field
gradients. When the single echo pulse sequence is replaced by a train of $n$%
\ pulses spaced apart by $t_{H}/n$, the signal at the time $t_{H}$, is
proportional to 
\begin{equation}
\left\{ \exp [-\frac{t_{H}}{n}/T_{2}-(\frac{t_{H}}{n}/\tau
_{D}^{{}})^{3}]\right\} ^{n}  \label{Carr}
\end{equation}
It is obvious that for large $n$\ this becomes a single exponential, being
the diffusion term eliminated by the train of pulses. One might think that a
similar argument could be used to explain why a train of pulses eliminates
the Gaussian factor in (\ref{Mpe}). However, one should understand the
conditions for the validity of (\ref{Carr}) to see whether equivalent
conditions apply to the REPE sequence. In order to have a contribution
affected by a train of pulses, its originating process must be affected.
This is independent of the particular decay law. \ The condition that
justifies the multiplication of echo amplitudes in (\ref{Carr}) is that at
each intermediate echo the initial conditions are recovered, although with a
lower amplitude. This requires that the missed amplitude must not retain any
correlation, classical or quantum, with the observed state. In the case of
molecular diffusion, which in principle might lead to classical
correlations, these are destroyed by the $\pi $-pulse because the phase
acquired from the field gradient accumulates excluding the backdiffusive
events from contributing to the observed polarization. Now let us go back to
the REPE sequence. On the basis of our experiment we have two categories of
processes. The ones characterized by $\tau _{\phi }$\ are not modified by
the pulse sequence and hence are not related to the variables affected by
these pulses. With respect to the other, we first observe that while
amplitude is modified by $n$\ it does not follow the product law. From this
last point we learn that correlations classical or quantum are present. The
pulse sequence is intended to modify dipolar spin dynamics. However, other
effects are simultaneous: Pulse imperfections and higher order terms of the
average Hamiltonian which are not inverted. A major part of the pulse
imperfections is accounted for by the renormalization of the attenuation
curves. While the remaining effects could, in principle, affect the PE
attenuation, reasonable magnitudes included in the numerical simulations in
small systems (see next section) cannot account, by themselves, for the
observed time scale of the decay. An effect present in the real experiments
is the unboundedness of the spin system. This huge space becomes available
to the spin correlations at a rate controlled by $n.$ Since they are
entangled with non inverted interactions, they could make it available to
irreversible processes. While the experiments, standing alone, cannot rule
out other phenomena, they lead us to tentatively attribute a major role to
the spin dynamics in the changing part of the attenuation. In order to get
an insight on how this mechanism emerges, we resort to the further hints
provided by numerical calculations on model systems.

\section{Numerical solutions}

The ZME sequence can only change the sign of the secular part of the
homonuclear dipolar Hamiltonian. Then, all the non-secular interactions are
uncontrolled processes that contribute to the attenuation of the PE. Here,
we will consider their effects calculating the exact evolution of the LP
during the defocusing and refocusing periods in a cyclopentadienil ring and
in a complete molecule of ferrocene with a single $^{13}$C.

Let us consider that the initial state, with only the $I_{1}$ spin
polarized, evolves during a time $t_{R}$ with the Hamiltonian 
\begin{equation}
{\cal H}_{{}}^{1}={\cal H}_{{}}^{y}+{\cal H}_{I}+{\cal H}_{n.s.}^{y}+{\cal H}%
_{IS}={\cal H}_{{}}^{y}+\Sigma ,  \label{ht1}
\end{equation}
where 
\begin{equation}
{\cal H}_{I}=-\hbar \gamma _{I}B_{1I}^{y}\sum_{k}I_{k}^{y}  \label{hi}
\end{equation}
and 
\begin{equation}
{\cal H}_{IS}=\sum_{k}b_{k}2I_{k}^{z}S_{{}}^{z}.  \label{his}
\end{equation}
Here, ${\cal H}^{y}$ (described by Eq. (\ref{Hii})) and ${\cal H}_{n.s.}^{y}$
are the secular and non-secular parts of the $I$-spin dipolar interaction
respectively. ${\cal H}_{IS}$ is the non-secular heteronuclear dipolar
Hamiltonian. Taking into account ideal $\left( \pi /2\right) $ pulses the
evolution continues for a time $t_{R}/2$ with $-[2]{\cal H}^{y}$. We
explicitly neglect the heteronuclear interactions during this period, since
it was verified that they do not produce any visible attenuation in the
observed time scale. In this way, the $S$-spin appears only in (\ref{his}),
and its effect can be taken into account by replacing ${\cal H}%
_{IS}\rightarrow \sum_{k}b_{k}I_{k}^{z}$. This substitution reduces to one
half the computations involved. Figure 6 shows the PE amplitude as a
function of $t_{R}$. The Gaussian fitting of the ferrocene experimental data
(Fig. 4) is included for comparison. Even when these numerical results
cannot reproduce the experimental data, they show that the non inverted
interactions produce a considerable destruction of the dipolar coherence
which manifests in the decay of $M_{PE}$. The dipolar coupling with the $%
^{13}$C spin is the most relevant term to produce attenuation in the
presence of spin dynamics, although in its absence, it does not produce a
significant decay, as it is shown by the oscillating curve in Fig. 6 where a
single proton is present. The calculations for a ring of five dipolar
coupled protons and for two rings show that the attenuation increases as
more spins are included. This trend implies that, in the highly connected
actual interaction network, a stronger attenuation, approaching the
experimental curve, should be expected. Conversely, the sequence of
solutions going up in Fig. 6, can be associated to a progressive reduction
of the dynamics obtained through the application of the REPE sequence. In
analogy with the inset in Fig. 4, it would reach the absence of decay when
dynamics is totally hindered ($n\rightarrow \infty )$. The lesson to be
drawn from this calculation is that a small non inverted term, although non
relaxing by itself, if mounted on a complex spin dynamics, it is amplified
and leads to strong dynamical irreversibility. Thus any reduction of the
complexity of the dipolar order (increase of $n$ in our experiment) will
decrease the amplification effect. While this slows down the polarization
echo decay, an asymptotic time scale is not reached. This is the case of the
ferrocene system.

Another kind of uncontrolled processes that could contribute to the PE
attenuation are those producing an exponential irreversibility. As a model
for these we can consider a fluctuating isotropic magnetic field. For
simplicity, we assume it is only effective during the evolution in the
rotating frame. Thus, the evolution of the density matrix during $t_{R}$ is
given by the quantum master equation \cite{NMR}: 
\begin{eqnarray}
\frac{{\rm d}\rho }{{\rm d}t} &=&-\frac{{\rm i}}{\hbar }\left[ {\cal H}%
_{{}}^{1},\rho \right] -\frac{1}{2\tau _{\phi }}\sum_{\alpha
=x,y,z}\sum_{j}\left[ I_{j}^{\alpha },\left[ I_{j}^{\alpha },\rho -\rho
(\infty )\right] \right]  \nonumber \\
&=&-\frac{{\rm i}}{\hbar }\left[ {\cal H}_{{}}^{1},\rho \right] -\widehat{%
\widehat{\Sigma }}(\rho -\rho (\infty )).  \label{master}
\end{eqnarray}
Notice that in the absence of dynamics (${\cal H}^{y}\equiv 0$) we would
obtain $\left\langle I_{1}^{y}\right\rangle \simeq \exp (-t_{R}/\tau _{\phi
})$. As in the previous calculation, the period of refocusing is governed by
the Hamiltonian $-[2]{\cal H}^{y}$. Fig. 7 shows the results for a single
cyclopentadienil ring. Initially the decay follows an exponential law with a
characteristic time $\tau _{\phi }$. However, after a while, it switches to
an exponential with a shorter characteristic time. The typical time for this
change is controlled by the dipolar dynamics. This can be clearly seen by
considering different time scales ($\hbar /d$) for the dipolar interaction
(Fig. 7). However, in our small systems, the decay rate of the new
exponential is not affected by this dynamics. Simulations on systems with
different sizes show that this rate increases with the number of spins
involved in the dynamics. Hence, we can expect that in unbounded systems,
where the number of correlated spins increases progressively, the decay
could not be represented by any finite number of switches. In that sense,
when the dipolar dynamics is strong as compared with the exponential process
($\tau _{mb}<$ $\tau _{\phi }$), the observed Gaussian tail can be thought
of as a continuous switch between progressively stronger exponential decays.
In the reciprocal case ($\tau _{\phi }<\tau _{mb}$), the experimentally
relevant decay occurs while the exponential process is dominant.

The lesson learned from these numerical results, is that there is an
emergent phenomenon in large many body systems through which spin dynamics
favors the irreversible processes. This phenomenon is the quantum analogous
to the one observed in a classical gas of rigid disks\cite{Bellemans}. The
many body interaction creates an instability that amplifies the numerical
rounding errors progressively limiting the efficiency of the time reversal.
Similarly, for a local spin excitation in a lattice, evolving according with
a cellular automaton dynamics\cite{Lacelle}, a single spin flip (not
included in this dynamics) prevents the full reversal of the polarization to
the original site producing its spreading.

\section{Conclusions}

We have studied the decay of the polarization echoes in different systems.
The ZME sequence in its original version, or even better in its modified
version providing a proper normalization, allows one to distinguish between
a tentative classification which recognizes two kinds of systems. They are
the dipolar dominated ones, which present a Gaussian component that can be
reduced by the REPE sequence (like ferrocene and cobaltocene), and those
with strong sources of relaxation where an exponential decay is already
manifested with the ZME sequence (cymantrene).

The REPE sequence, designed to reduce the time scale of the spin dynamics by
increasing the number $n$ of pulses, is able to give further information on
the first group by revealing the existence of underlying relaxation
mechanisms for large enough $n$'s as it is the case of cobaltocene. Looking
for an experimental test to rule out possible experimental artifacts in
these results, one could resort to alternative pulse sequences which also
achieve the dynamics reversal with less demanding train pulses\cite
{Suter,EPER}. Further tests of the sensitivity of the REPE sequence as a
tool to characterize new materials are experiments in which a dipolar
dominated system is progressively doped with local centers of relaxation
(e.g. a solid solution of cobaltocene in ferrocene). Besides, the REPE
sequence can be readily adapted to be used under MAS by rotor synchronizing
the pulses according to the prescriptions of ref.\cite{Tomaselli}.

Because of the deep physical meaning of the Polarization Echo, beyond its
potential applications as a practical tool, the ZME and REPE\ sequences open
an exciting new field of study in fundamental physics. It is the onset of
irreversibility in interacting many body quantum systems. This is a
particular area of the growing discipline of quantum chaos\cite{QC} whose
theoretical development is still in its infancy. On one side the magnitude
of the polarization echo coincides with the most convincing signature of
quantum chaos\cite{Peres}: the exponentially fast decrease of the overlap
between states evolved, from the same initial condition with slightly
different Hamiltonians ${\cal H}$ and ${\cal H}+\Sigma $. The connection is
evident for a system of $N$ interacting spins since we can write $%
M_{PE}=2^{N}\,\,{\rm Tr}\left( {\bf \rho }_{{\cal H}}{\bf \rho }_{{\cal H+}%
\Sigma }\right) -1$, which is a way to evaluate this overlap. This
expression has close resemblance to some definitions of entropy\cite{Entropy}%
. Independently, -$\ln [M_{PE}(t_{R})]$ is a measure of how the inefficiency
in our dynamics reversal increases with time. When the polarization is a
conserved magnitude within the observed time scale, $M_{PE}(t_{R})$ is a
measure of the inverse of the volume $\Omega _{t_{R}}$ occupied by the
polarization at the PE time: -$\ln [M_{PE}(t_{R})]\propto \ln [\Omega
_{t_{R}}]$ which gives an ``entropic'' meaning for both the overlap and the
polarization echo. To further visualize this, one might assume that after a
few pulses the shape of the distribution function for the refocused
polarization (e.g. $P(x,t)=1/\Omega _{t}$ for $\left| x\right| <\Omega
_{t}/2 $ where $t$ is an echo time) remains the same (i.e. the echo would
represent a scale transformation). In this case $-\int P(x)\ln [P(x)]{\rm d}%
x $ grows proportionally to $-\ln [P(0,t_{R})]=-\ln [M_{PE}(t_{R})]$.

Thus, the Gaussian regime for the attenuation of the PE in our experiments
seems to indicate an anomalous decrease (growth) of the overlap
(``entropy'') not predicted within any of the present one body theories. Our
experiments might be interpreted as an indication that {\it infinite}
interacting many body quantum systems present an extreme intrinsic
instability towards dynamical irreversibility. The experiments, even with
imperfections, are much ahead of the possibilities of the current theory and
they have the leading word.

\section{Acknowledgments}

The authors express their gratitude to Professor E. Hahn, Professor R. R.
Ernst, Professor B. Meier, Professor A. Pines and Professor S. Lacelle for
stimulating discussions at different stages of this work. This work was
performed at LANAIS de RMN (UNC-CONICET), where the ferrocene single crystal
was grown by Rodrigo A. Iglesias. Financial support was received from
Fundaci\'{o}n Antorchas, CONICET, FoNCyT, CONICOR and SeCyT-UNC. The authors
are affiliated to CONICET.

\newpage

{\bf Figure 1}: {\bf a) }ZME{\bf \ }sequence for refocusing the dipolar
evolution of the $^{1}$H spin polarization in the laboratory frame (see
text). The solid and dashed lines show the main and secondary amplitude
polarization pathways {\bf b)} Alternative version of the ZME sequence where
the refocusing occurs in the rotating frame. The polarization maximum at $%
t_{2}=t_{m}$ provides an experimental point for data normalization. It
allows a straightforward comparison of the PE attenuation in different
systems. {\bf c}) Reduced Evolution Polarization Echo (REPE) sequence to
control the dipolar spin dynamics. For a given time $t_{R},$ setting $%
t_{1}=t_{R}/n$ and $t_{2}=\left[ \frac{1}{2}\right] t_{1}$, the development
of multiple spin order is reduced by increasing $n$.

\smallskip
{\bf Figure 2}: Normalized polarization echo amplitude as a function of $t_R$
for polycrystalline samples of cymantrene, ferrocene and cobaltocene, using
the ZME sequence. The lines are fittings to a Gaussian (ferrocene),
exponential (cymantrene) and a product of both functions (cobaltocene).

\smallskip
{\bf Figure 3}: $^{13}$C NMR spectrum of a ferrocene single crystal using
the REPE sequence with: $n=1$, $\omega _{1I}/2\pi =62$ {\rm kHz}, $t_C=2$
$m{\rm s}$, $t_S=1$ $m{\rm s}$, $%
 t_d=53$ $\mu {\rm s}$ and $t_R=8$ $\mu
{\rm s}$. The two peaks indicate the
 presence of two magnetically
non-equivalent sites. The on resonance peak
 corresponds to molecules with its
fivefold molecular symmetry axis at
 approximately 20$^{\circ }$ with respect
to the external magnetic field.

\smallskip
 
{\bf Figure 4}: Attenuation of the polarization echo in the ferrocene single
crystal as a function of $t_{R}$ for $n=1$ (REPE sequence). The line
represents a Gaussian fitting with the characteristic time $\tau
_{mb}^{1}=(245\pm 5)$ $\mu {\rm s}$ as the only free parameter. {\bf Inset}:
PE attenuation for progressively reduced dipolar dynamics, $n=1,2,8,16$. No
asymptotic regime is reached within the experimental time scale. The solid
lines are Gaussian fittings yielding characteristic times $\tau
_{mb}^{2}=(335\pm 10)\,\mu {\rm s,}\tau _{mb}^{8}=(650\pm 20)\,\mu {\rm s}\,$
and $\tau _{mb}^{16}=(1070\pm 60)\,\mu {\rm s}$.

\smallskip
{\bf Figure 5}: Attenuation of the polarization echo in the cobaltocene
sample as a function of $t_{R}$ for $n=1,2,5,8,16$ with $\omega _{1I}/2\pi
=56$ {\rm kHz}, $t_{C}=85$ $\mu {\rm  s}$, $t_{S}=150$ $\mu {\rm s}$, and
$t_{d}=85$ $\mu {\rm s}$. The
 experimental data show a clear crossover
between a dominant Gaussian
 attenuation to an exponential one. The solid
lines represent fittings of the
 whole set of data to equation (\ref{Mpe})
yielding $a=\left( 130\pm
 10\right) \,\mu {\rm s}$, $b=\left( 90\pm%
10\right) \,\mu {\rm s}$ and $\tau
 _{\phi }^{{}}=\left( 640\pm 20\right)%
\,\mu {\rm s}$.

\smallskip
{\bf Figure 6}: Numerical calculation of the attenuation of the polarization
echoes considering the effects of the non-inverted interactions. The thick
solid and dashed lines correspond to a cyclopentadienil ring and to a
ferrocene molecule respectively. The Gaussian curve is the fitting of the
ferrocene experimental data of Fig. 4. The attenuation is mainly caused by
the destruction of the dipolar coherence induced by the heteronuclear
dipolar coupling. ${\cal H}_{IS}$ does not produce relaxation in the absence
of homonuclear dipolar interactions (oscillating curve).

\smallskip
{\bf Figure 7}: Attenuation of the polarization echoes for a
cyclopentadienil ring considering an exponential irreversible process (See
Eq. (\ref{master})). The different curves represent a progressive expansion
of the dipolar time scale ($\hbar /d$). Here $d_1=2610\ {\rm Hz}\times 2\pi
\hbar $ is the nearest neighbor dipolar coupling. The initial decay rate, $%
1/\tau _\phi =500\,{\rm Hz}$, switch to a stronger one after a while. The
upper curve correspond to the exponential decay, $\exp (-t_R/\tau _\phi )$,
in the absence of dipolar dynamics.

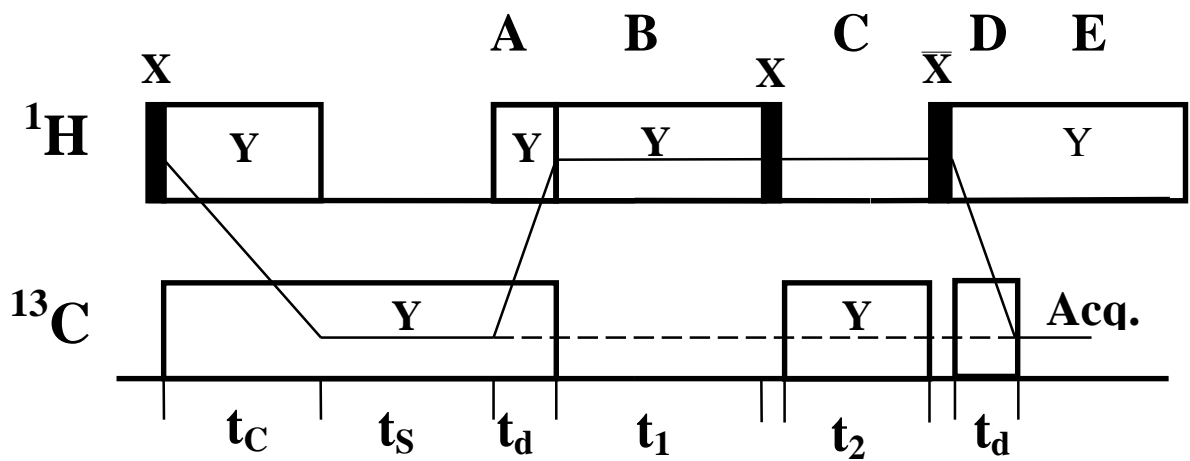
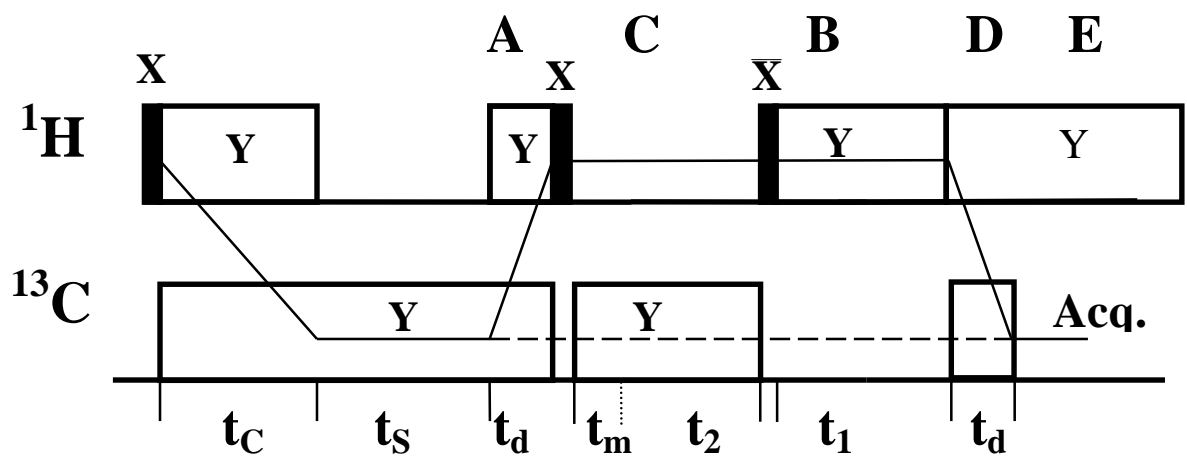
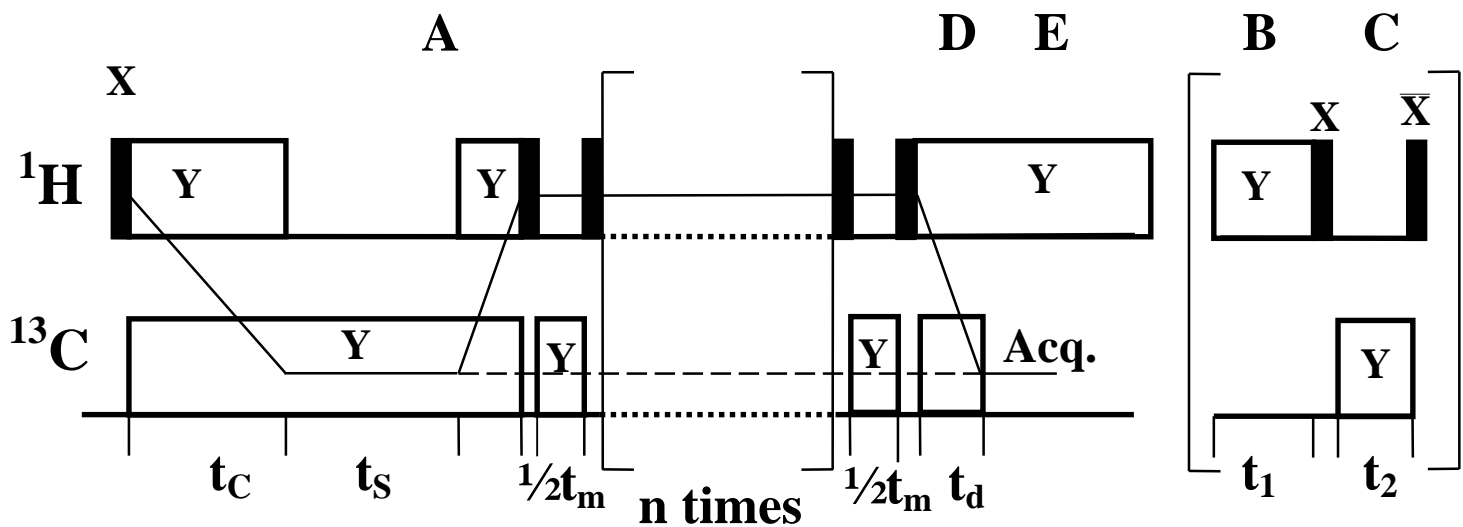
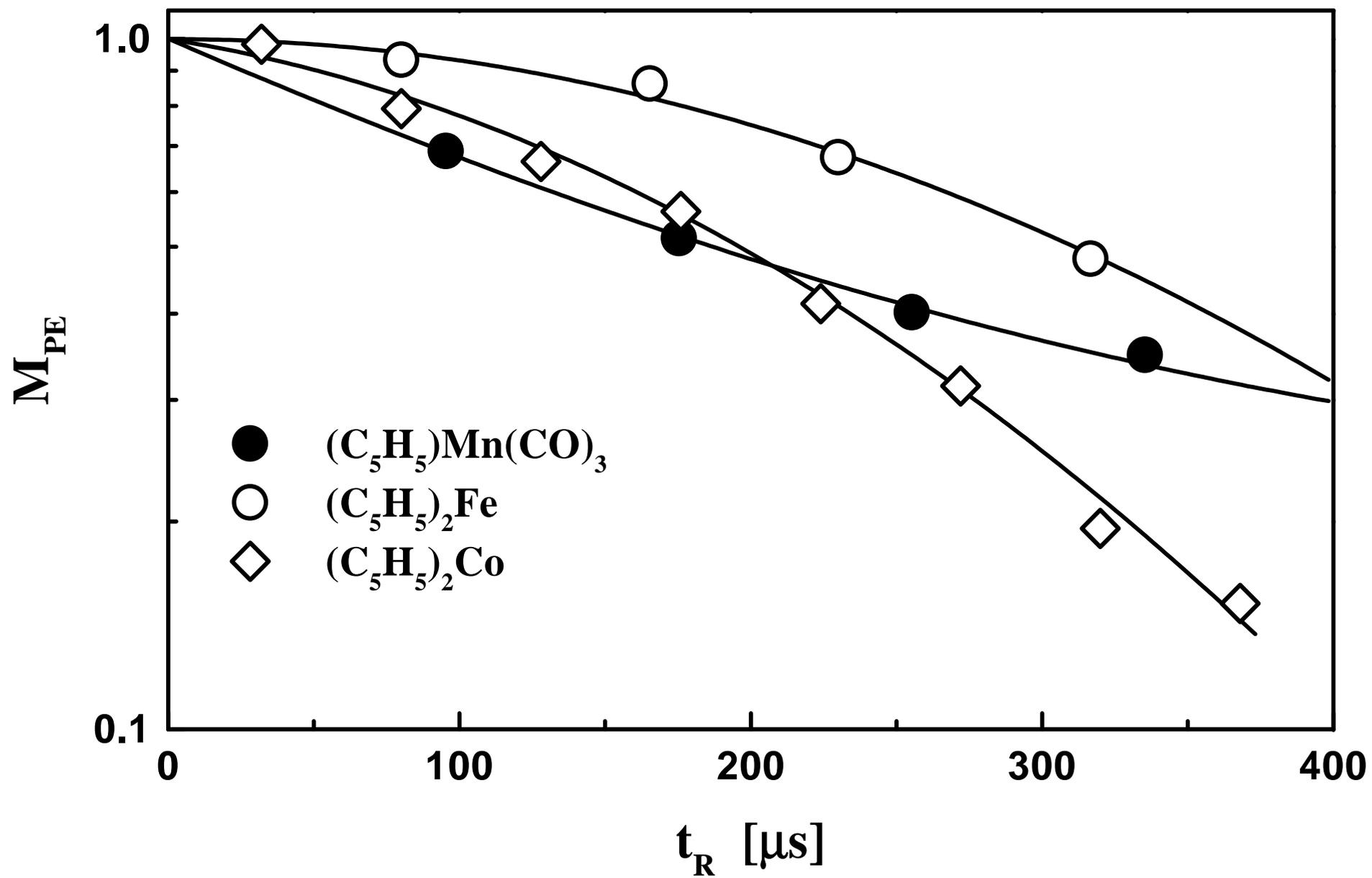
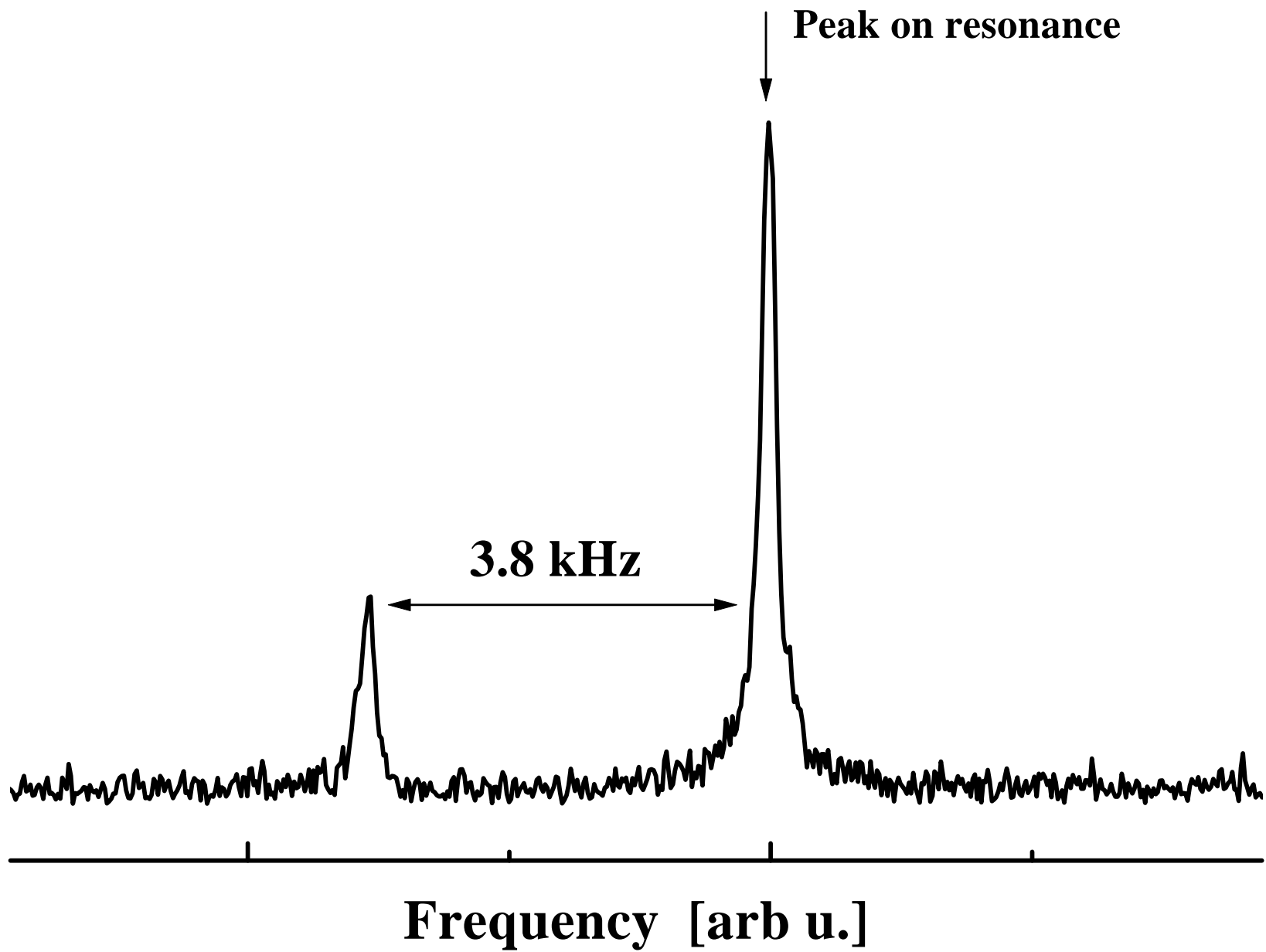
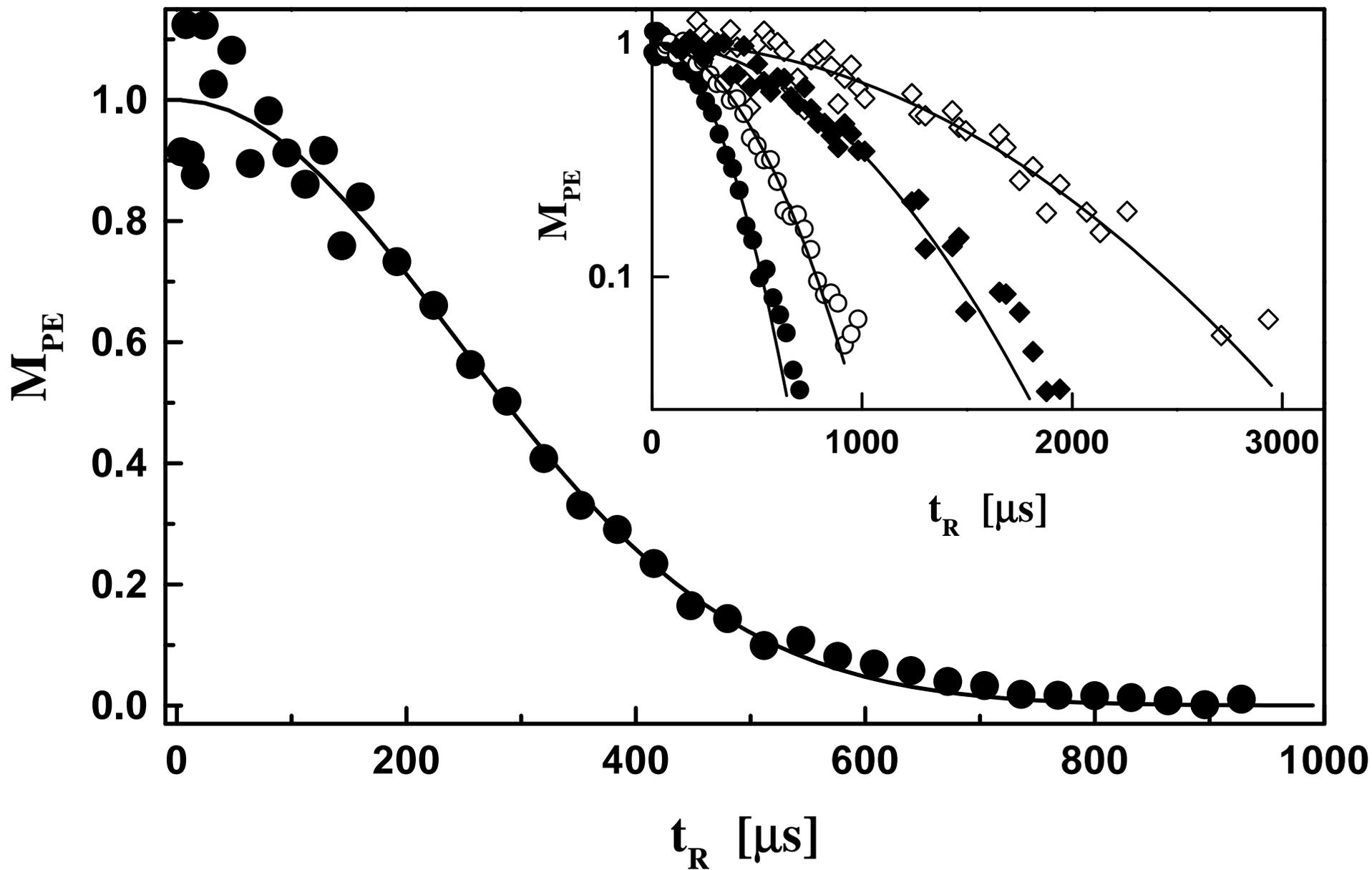
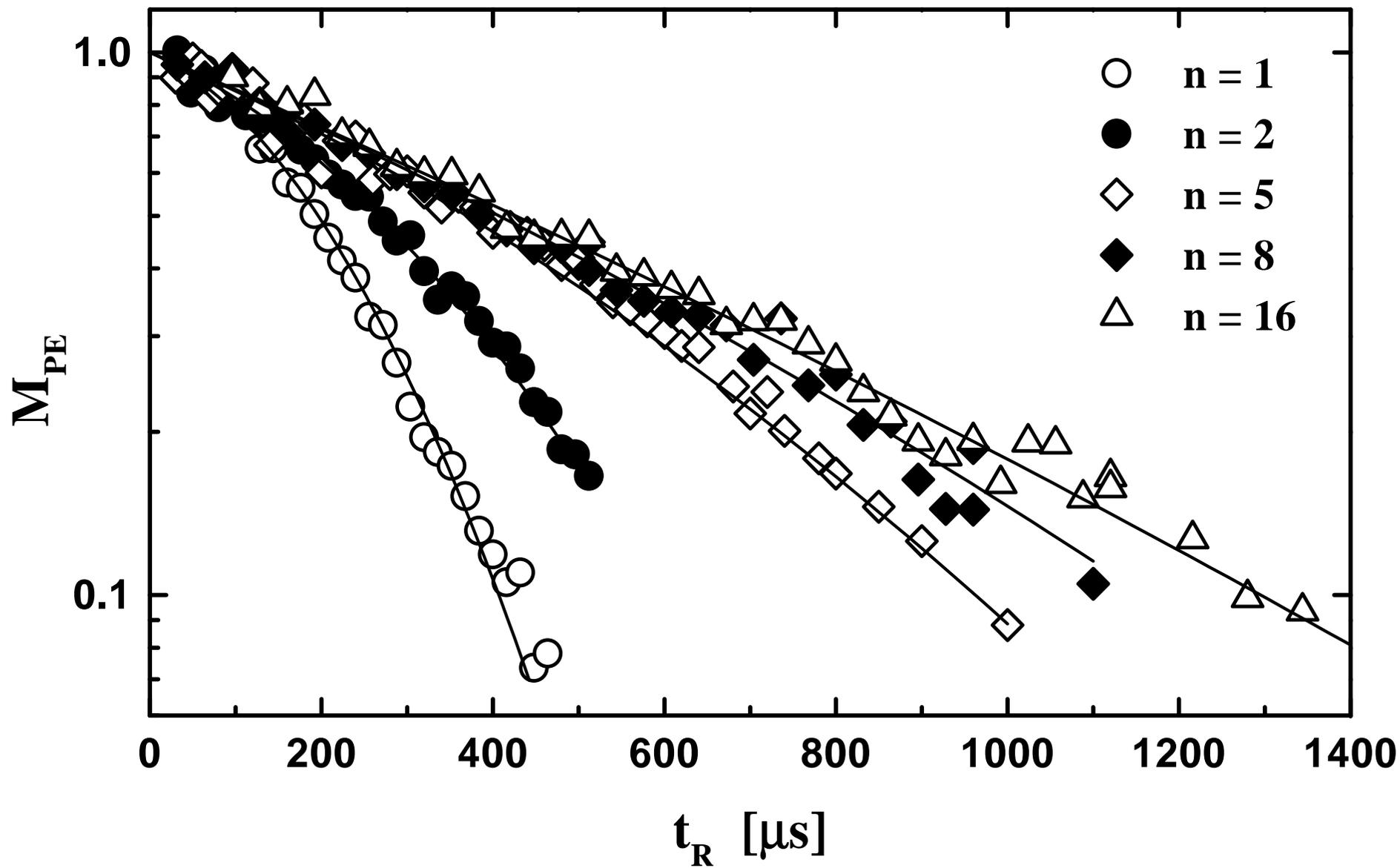
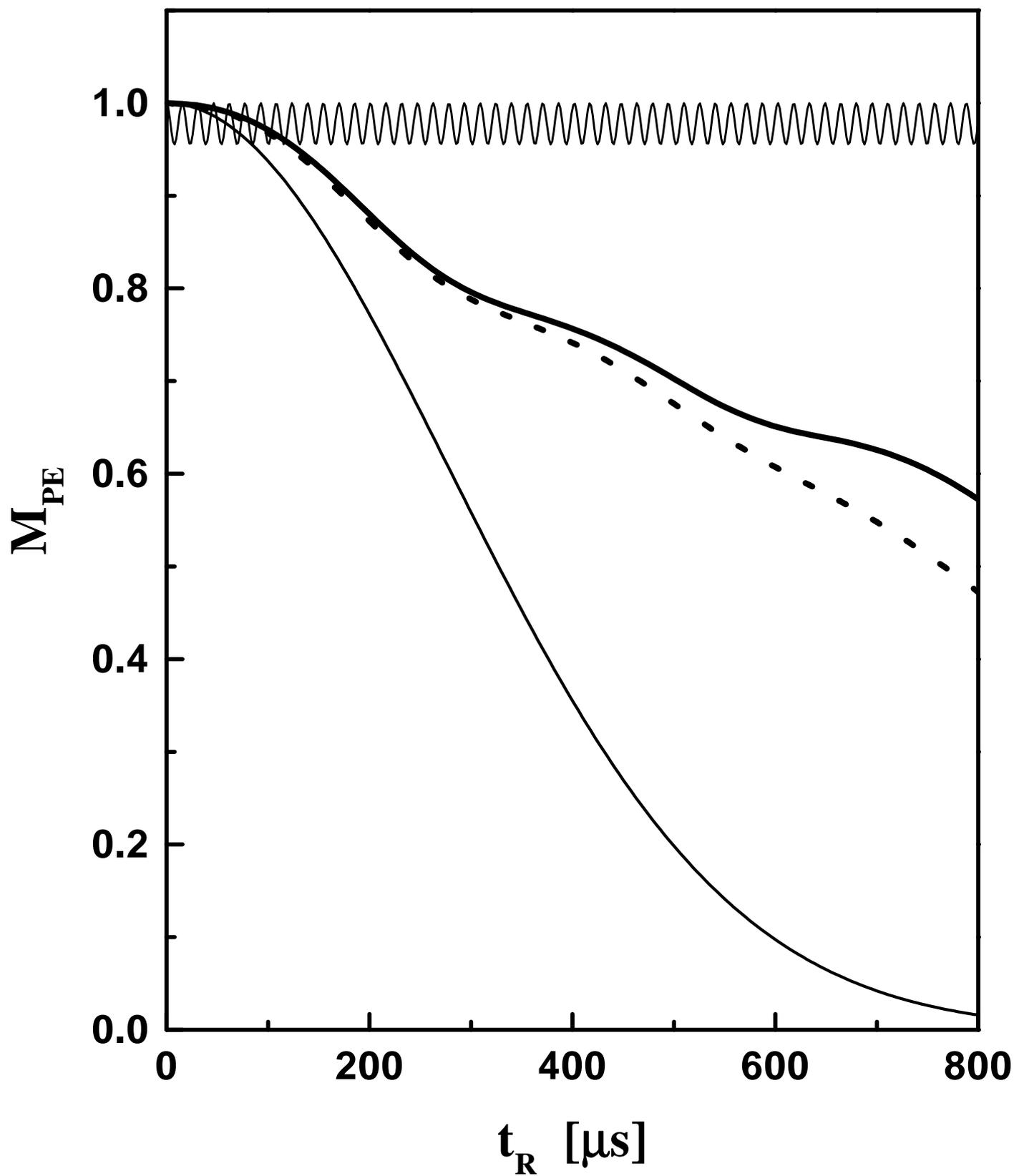
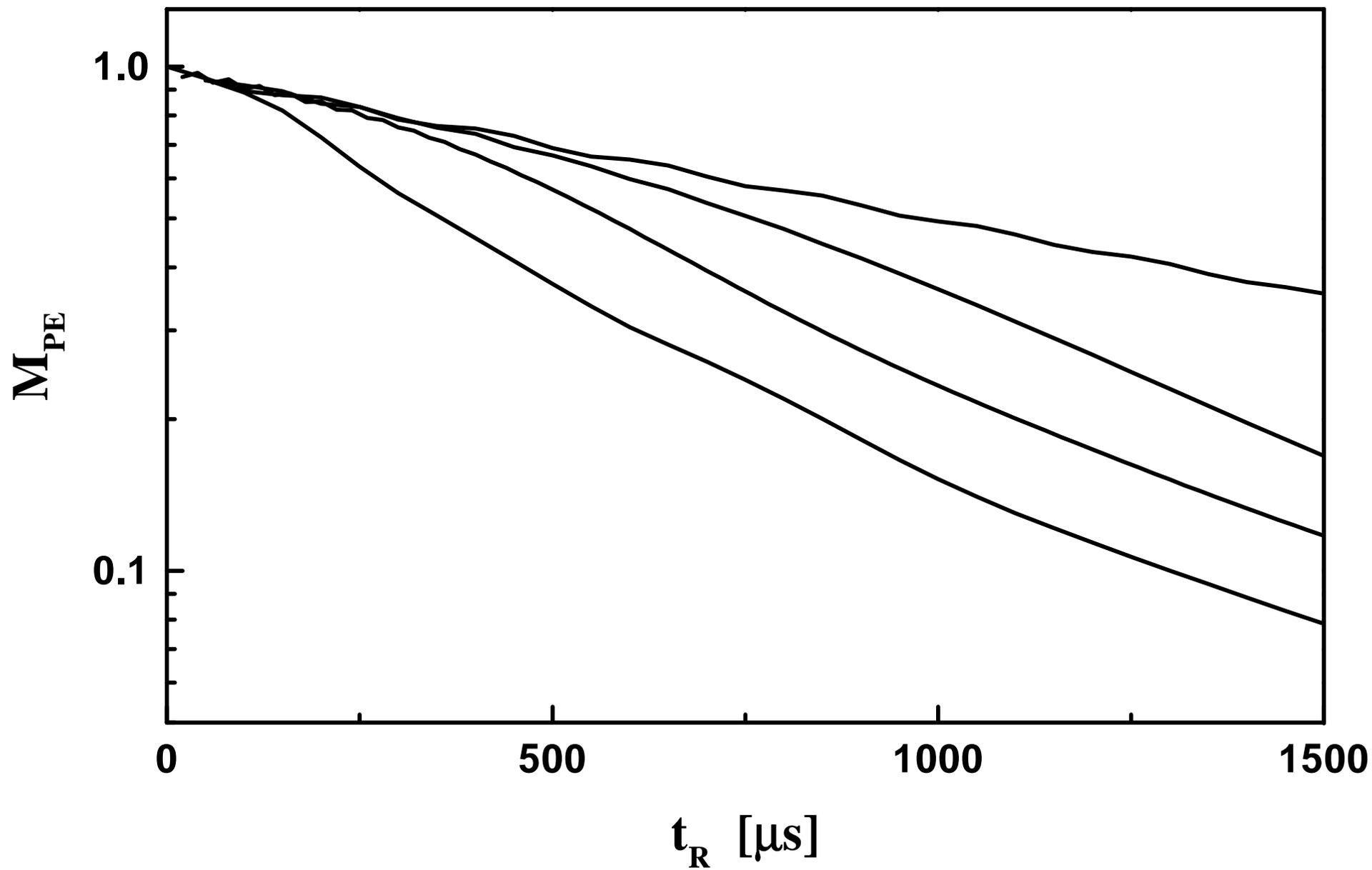
\end{document}